\documentstyle[epsf]{mn}

\title[Superiority of the Minimal Spanning Tree]
{The Superiority of the Minimal Spanning Tree in Percolation 
Analyses of Cosmological Datasets}

\author[S.P. Bhavsar \& R.J. Splinter]
{Suketu P. Bhavsar$^{1}$ and Randall J. Splinter$^{1,2}$ \\
$^{1}$Department of Physics \& Astronomy,
University of Kentucky, Lexington, KY 40506-0055 \\
$^{2}$Center for Computational Sciences, 
University of Kentucky, Lexington, Kentucky 40506-0045}
\date{Accepted 1996 May 24; in original form 1996 February 9}

\begin{document}
\maketitle

\begin{abstract}

In this work  we demonstrate the  ability of the Minimal Spanning Tree
to duplicate the  information contained within  a percolation analysis
for  a  point dataset.   We  show how   to  construct the  percolation
properties from the Minimal Spanning Tree, finding roughly an order of
magnitude improvement in the  computer time required.  We apply  these
statistics to   Particle-Mesh simulations   of  large-scale  structure
formation.  We consider  purely scale-free Gaussian initial conditions
($P(k) \propto   k^n$, with $n =   -2, -1, 0  \ \&  +1$) in a critical
density universe.  We find  in  general the  mass of   the percolating
cluster is a  much  better quantity by  which  to judge  the  onset of
percolation than the length of the percolating cluster.

\end{abstract}

\section{Introduction}

In  1982  Zel'dovich  suggested that   the  statistics derived  from a
percolation analysis of  the density distribution  might be  useful in
characterizing  the  topology of  the  distribution.  Soon  after that
Shandarin  (1983) and  Shandarin   \& Zel'dovich (1983)  explored  the
possibility that the percolation properties of the galaxy distribution
might  provide a   useful  measure of  the  topology  of  the observed
large-scale structure and act as  a method for discriminating  between
various cosmological   models.   Einasto,  et al.    (1984)    applied
percolation  analysis to the  CfA  I catalog. Their findings indicated
that the  large-scale distribution of galaxies  was  consistent with a
network-like  structure.  Bhavsar    \&   Barrow (1983)   applied  the
percolation  method  to theoretical   studies of $N$-body  models with
power law initial conditions.   In a $\Omega =  1$ universe they found
that the  $n=-1$  case agreed much  better with  observations than the
$n=0$ case.   Additional  work which centered on   the CDM spectrum by
Melott \& Shandarin (1983) and Davis, et al.  (1985) demonstrated that
the CDM model also has  a connected, network-like structure as opposed
to a clumpy distribution.  Dekel  \& West (1985)  pointed out that the
percolation  method would depend strongly  on  the mean density of the
sample,  which  would  make the  method   difficult to use for  sparse
datasets. Recent  work  by Yess \&  Shandarin  (1995) has demonstrated
that a percolation analysis of a continuous density field on a lattice
is  able to  provide  robust  statistical measures  of  the underlying
distribution which do not suffer from the  earlier criticisms of Dekel
\& West (1985).

In their  search for  an  objective method  for the  identification of
filaments in observational  datasets Barrow, Bhavsar \&  Sonoda (1985)
introduced   the Minimal  Spanning  Tree  (MST)  into the cosmological
literature.  The MST  is a graph theoretical  construct which has been
used to  quantify  patterns in  datasets (Zahn  1971).  Barrow,  et al
(1985) developed several statistics based upon the MST from which they
were able to  differentiate  between a Poisson distribution  of points
and   several  observational    datasets.   The  introduction   of   a
bootstrap-based method,  referred to as ``shuffling'', allowed Bhavsar
\& Ling (1988) to ascertain the existence of  the filaments in the CfA
survey as real objects and not visual artifacts.  Recently Krzewina \&
Saslaw (1995) have introduced several additional statistics based upon
the MST which    they use to compare   a  subset of  the  Southern Sky
Redshift  Catalog  (SSRC) to an   $N$-body   simulation and a  Poisson
distribution.

It is possible to construct the MST  for any distribution of points in
space (Gower \& Ross 1969; Abraham 1962; Zahn 1971).  The MST uniquely
connects a set  of  $N$ points (referred to  as  ``nodes'') with $N-1$
lines (referred to as ``edges'') in such a  way as to minimize the sum
of the $N-1$  edges. Consequently   closed  paths are excluded.   This
property has  been exploited  in  the  past as  a  way to  objectively
identify  filamentary features(Bhavsar  \& Ling  1988).   The skeletal
pattern defined by the  MST  can then be  used  to define a  number of
objective statistics (Barrow, et al. 1985; Krzewina \& Saslaw 1995) 
which describe the clustering of the data points.

In this work  we first demonstrate that for  a point data  set the MST
contains  all of    the   information which   is contained  within   a
percolation  analysis  for  that dataset.    We then  demonstrate  the
relative robustness of  various percolation based statistical measures
of the clustering for a Poisson dataset.  We should stress that rather
than emphasizing a single number,  such as the percolation  threshold,
we base our analysis on  curves derived from the percolation analysis.
We work  with point datasets  as the original percolation studies did.
Thus we use the simulations ``as they are'' and  the techniques can be
applied directly  to the positional data  from  galaxy catalogs.  This
avoids problems with  boundary conditions at the  edge of  the sample,
and  determining a density  field  from observational data.  The  time
efficiency  obtained   using the MST  to   investigate the percolation
properties has  encouraged us to apply the  statistics  to a series of
large $N$-body simulations.  These  studies should, we hope,  pave the
way for the eventual analysis for data from the large redshift surveys
currently underway.

\section{Percolation as a Subset of the MST}

To    build the MST we    use  Prim's algorithm   (1957). The simplest
algorithm to construct explicitly the  MST of a graph, $\Gamma$, first
picks an arbitrary node  of $\Gamma$ and then  adds the connected edge
of smallest length.  This edge and the two  nodes at its ends form the
partial tree,  $\Pi_1$. The $k$th partial  tree, $\Pi_k$, is formed by
adding to $\Pi_{k-1}$ the shortest edge  connecting $\Pi_{k-1}$ to any
nodes of $\Gamma$ not already in $\Pi_{k-1}$. If $\Gamma$ contains $n$
nodes  then  $\Pi_{k-1}$  is  the  required  MST. Therefore,  there is
clearly small-scale   information in the  tree  because of the  way in
which it is  built, but the MST  also contains large-scale information
because the sum of all the edge lengths is a minimum.   Once an MST is
constructed, separation is the  operation of removing all  edges whose
length exceeds some cutoff.

The percolation method we  use was discussed  in detail in Bhavsar  \&
Barrow (1983). The method consists of enclosing individual data points
by a sphere of radius  $R$ centered on the   data point.  All  spheres
which intersect form a cluster. Typically a distribution of points and
their enclosing sphere's is charaterized by some critical value of $R$
at which the  length of the longest  single connected  chain of linked
spheres grows to of order the size of the  system. If this occurs then
the system is said to percolate (Hammersley \& Welsh 1980).

Now consider the following short  thought experiment.  Assume that the
data  set  has just  percolated,  so that   the radius  of the spheres
surrounding  each  data   point   is given   by  percolation threshold
$l_{perc}$.  The distance between  the  two most spatially   separated
points in any cluster will be $2 \times  l_{perc}$.  Therefore, if the
MST for the same dataset is separated using  a separation length of $2
\times l_{perc}$,  subtrees will be  identified which are separated by
at least  $2 \times l_{perc}$.   As a consequence  if we build the MST
and begin separating the MST we should find that  the linear extent of
the largest sub-tree should  exhibit exactly the  same behavior as the
longest percolating cluster determined  by a percolation analysis.  In
fact  carrying through the  thought  experiment for  a series of  edge
lengths, we  conjecture that  separating  the MST at every  successive
edge length starting from the largest to  the smallest edge length, we
recreate  the  entire percolation  analysis   at every possible sphere
radius.  Since this is accomplished   by just one construction of  the
MST and  subsequent separating, the   saving in computational  time is
enormous.  Our conjecture  has been verified by numerical  experiments
which follow.

The growth of the percolating cluster as a  function of the separating
length and also the  sphere radius is shown  and compared in figure  1
for a Poisson distribution of  $32^3$ particles.  This plot shows only
one such dataset.     We have tested  this  method  using  many random
realizations and consistently   find  the same  result.  To  make  the
comparison between the  two  curves more  qualitative  we compute  the
$L_1$ error which we define as

\begin{equation}
L_1 = {1 \over N} \sum_{i=1}^N \mid l_i^{percolation} - l_i^{MST} \mid, 
\end{equation}

\noindent where $l_i^{percolation}$ is the  length of the  percolating
cluster determined using the percolation code,  and $l_i^{MST}$ is the
length of the longest cluster using the MST/Separation method proposed
here. For figure 1 we find $L_1 = 1.2 \times  10^{-9}$, clearly at the
round-off level. This result is typical for the method.

The percolation method scales as  $O(N^2)$ for {\it each} radius  $R$.
So to identify the percolation threshold requires a significant amount
of  computer   time.  Though building   the MST   is also  a  $O(N^2)$
algorithm  the   separation process    requires  significantly   fewer
operations.  As  a consequence,  percolation analysis required roughly
46.5 CPU hours to produce the percolation  graph in figure 1,  whereas
the MST/separation  method required only  4.8 CPU  hours on a  Silicon
Graphics Indigo2 to produce the  identical plot (also shown in  figure
1)!   A savings  of roughly  an order  of  magnitude  in runtime. This
saving can be crucial depending on the size of the dataset.

\begin{figure}[t]
     \epsfxsize = 3.0truein
     \hskip 1.5truein
     \epsfbox{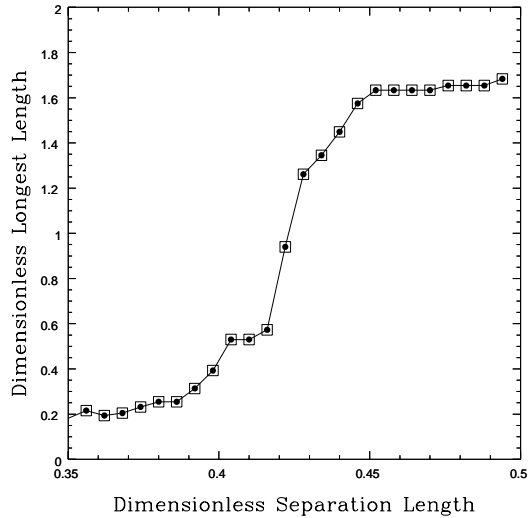}
     \caption{Percolation of a Poisson distribution using the 
              percolation code of Bhavsar \& Barrow (1983) in 
              open boxes and the MST based algorithm presented
              here in filled circles.
              The dimensionless separation length is defined as
              $l/n^{-1/3}$, where $n$ is the particle density.} 
\end{figure}

\section{Percolation Statistics for a Point Dataset}

In the past only the linear extent of the percolating cluster has been
considered a primary statistic  (Bhavsar \&  Barrow 1983).  In  recent
years Shandarin and his collaborators  (Klypin \& Shandarin 1993; Yess
\& Shandarin 1995) have extended  the percolation method to continuous
density fields  on a lattice and  demonstrated the  robustness of such
methods for studying  the large-scale  distribution  of mass. 

\onecolumn

\begin{figure}
     \epsfxsize = 7.0truein
     \hskip .5truein
     \epsfbox{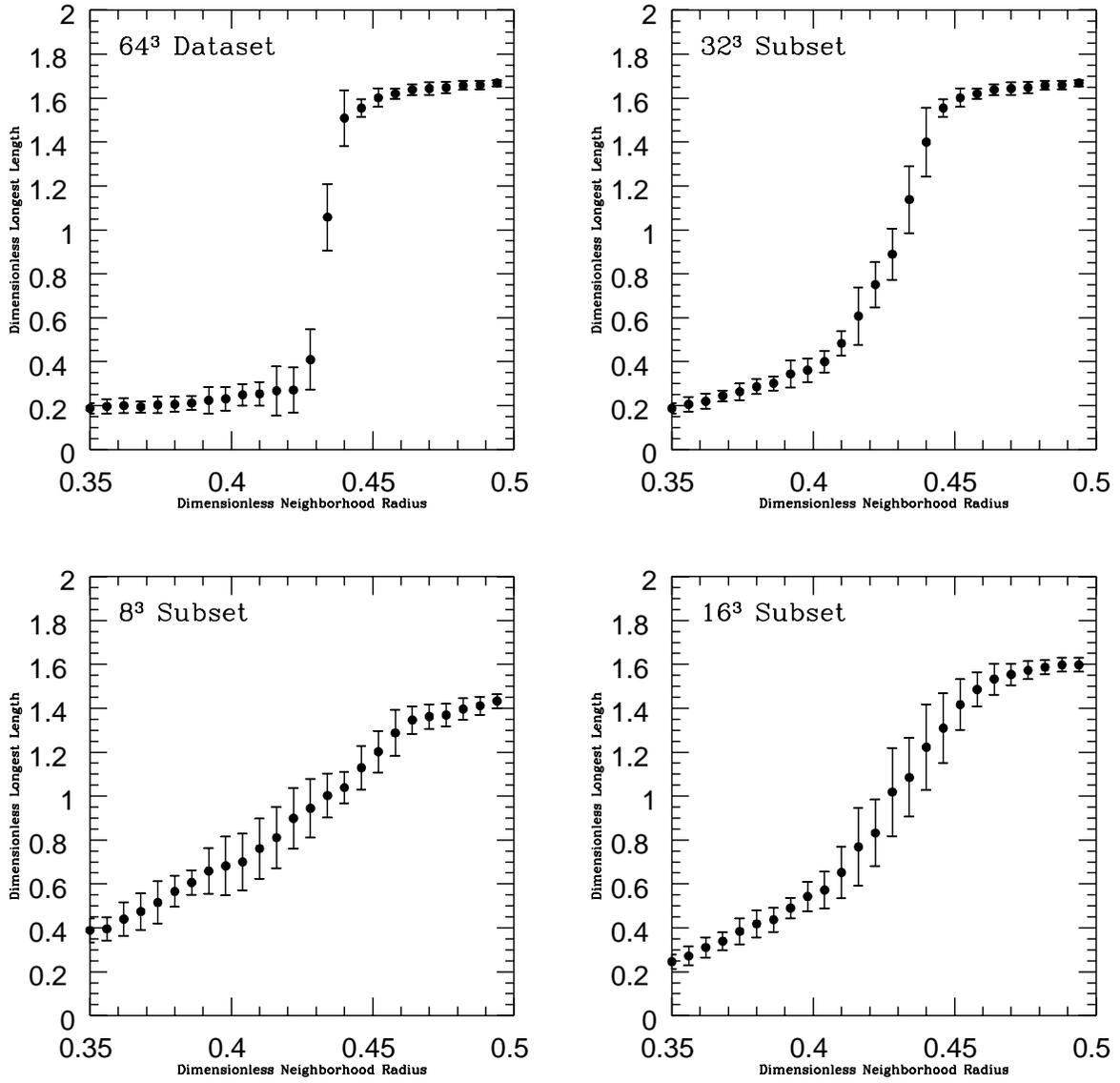}
     \caption{Robustness test for the linear extent of the percolating 
              cluster for a Poisson distribution of particles.
              The upper left plot is the entire $64^3$
              dataset. The upper right is a $32^3$ subset, the lower
              right is a $16^3$ subset, and the lower left is a $8^3$
              subset.
              The dimensionless neighborhood radius is defined as
              $l/n^{-1/3}$, where $n$ is the particle density.} 
\end{figure}

\clearpage
\onecolumn

\begin{figure}
     \epsfxsize = 7.0truein
     \hskip .5truein
     \epsfbox{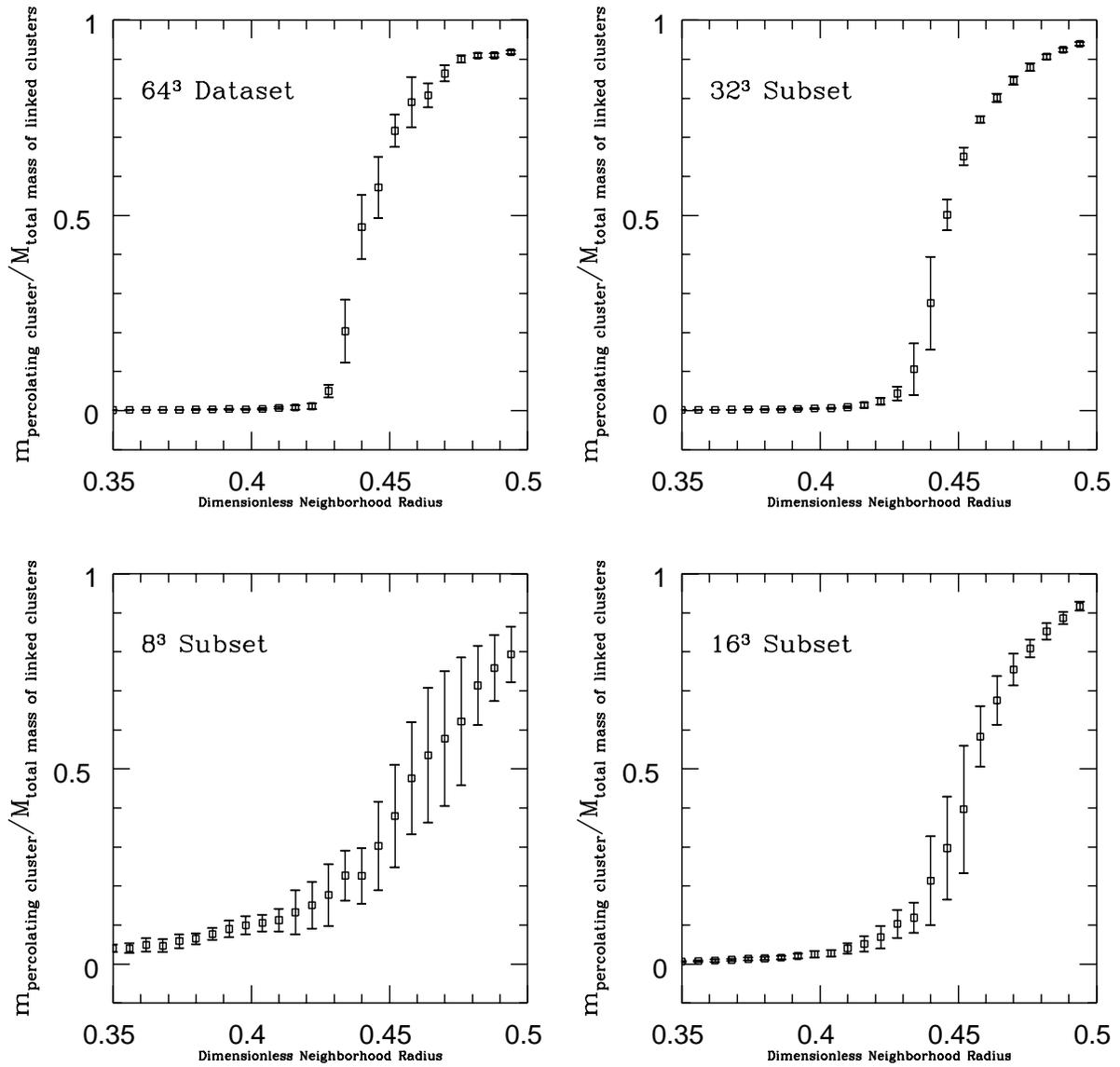}
     \caption{Robustness test for the total mass of the percolating
              cluster for a Poisson distribution of particles.
              The upper left plot is the entire $64^3$
              dataset. The upper right is a $32^3$ subset, the lower
              right is a $16^3$ subset, and the lower left is a $8^3$
              subset.
              The dimensionless neighborhood radius is defined as
              $l/n^{-1/3}$, where $n$ is the particle density.} 
\end{figure}

%\begin{figure}
%    \epsfxsize = 7.0truein
%    \hskip .5truein
%    \epsfbox{./maxmass.ps}
%    \caption{Robustness test for the ratio of the mass of the percolating
%             cluster to  the total mass of the linked clusters. 
%             The upper left plot is the entire $64^3$
%             dataset. The upper right is a $32^3$ subset, the lower
%             right is a $16^3$ subset, and the lower left is a $8^3$
%             subset.}
%\end{figure}
\twocolumn

\noindent Here  we wish  to present a new set  of statistics based upon 
percolation using the MST based algorithm for point datasets.

The  first statistic we  present  for comparison  is the  usual linear
extent  of the percolating cluster  as  a function of the neighborhood
radius $R$, the radius  of the  spheres surrounding  each  point.  The
second is the mass of the  percolating cluster normalized by the total
mass in the  simulation as a function of  the neighborhood radius. To
test the robustness of each of these statistics we generate 10 Poisson
distributions varying the number of  particles  in the box.  Figure  2
shows the linear extent  of the percolating  cluster for four particle
densities.  The first is $64^3$ particles in a box of size $64^3$, the
second is a $32^3$ subset of the original $64^3$ particles in the same
volume, the third is  a $16^3$ subset and  the final is a $8^3$ subset
of the original $64^3$ particles.

\noindent Figure 3  shows  the mass  of  the percolating cluster  as a
fraction of the total mass in the simulation for the same four subsets
of  particles.  Each point  is the average  over the ten realizations,
and  the  error  bars  represent the   $1\sigma$  deviations  from the
averages.
%Figure 4 shows  the ratio of the mass of
%the percolating cluster to the total mass of  the linked clusters as a
%function of the neighborhood radius.

Interestingly the mass of the percolating cluster appears to be a more
robust indicator of the onset of percolation than the linear extent of
the cluster. This isn't unexpected. As the particle density is reduced
shot noise due to undersampling can have a much more serious impact on
the length of the cluster than its mass. For instance, by removing one
particle its possible that the length of the percolating cluster could
change  dramatically, but  it is unlikely  that  removing one particle
will have much of an effect on the total mass of the cluster. Further,
the mass curve (Figure  3) demonstrates that it is  a much more robust
measure of   the percolation properties  of  the dataset.   The curves
allow one to accurately estimate  the percolation length for as little
as 1/64th  of the original particle  density, which corresponds to the
$16^3$ subset.  Even for the $8^3$  subset where there  is a factor of
512 fewer particles the percolation  length can be estimated to within
20\% or so.  Based  upon the linear extent  of the cluster both of the
$8^3$ subset and   the   $16^3$ subset are  relatively   worthless  in
estimating the percolation threshold.

\section{N-Body Methods}

The Particle-Mesh (PM) code  used to generate  the simulations used in
this work has been described in detail by Melott (1986). The code is a
standard  PM code, except that  it  uses  a  staggered grid to  obtain
slightly better force resolution (Melott et al. 1988). The simulations
use $128^3$ particles on a  comoving $128^3$ mesh. For the percolation
studies  here we use  a  $32^3$  subset  of  those particles.   We ran
simulations for four different  power law initial spectra,  $n = 1, 0,
-1, -2$, all for a $\Omega = 1$ universe.  Ten realizations of each of
the above four spectra were performed. These realizations were studied
at the nonlinear wavenumbers  $k_{nl} = 32, 16,  8, 4$ and the initial
conditions; $k_{nl}$ is defined by

\begin{equation}
\sigma^2 = a^2 \int^{k_{nl}}_0 P(k) d^3k = 1,
\end{equation}

\noindent where $P(k)$ is  the initial power  spectrum of the  density
fluctuations, and $a$ is the cosmic expansion factor.

The percolation  statistics were run on  the ten  realizations at each
evolutionary  stage.  Then the averages  and the 1 $\sigma$ deviations
were computed.  The results for the percolation statistics are plotted
in figure 4. To  make the comparisons more  qualitative we compute the
Kolmogorov-Smirnov (KS) statistic and significance level (Press, et al
1992) between each   of the curves   plotted  in figure  4. These  are
presented in tables 1 \& 2.

\begin{table}
      \caption{Significance levels computed between the 4 $N$-body
               models for the length of the percolating cluster 
               statistic.
              The dimensionless neighborhood radius is defined as
              $l/n^{-1/3}$, where $n$ is the particle density.} 
\begin{tabular}{|c||cccc|}

Spectral Index & -2 & -1 & 0 & 1 \\ 
-2 & 1.0 & 3.03(10)$^{-3}$ & 1.31(10)$^{-2}$ & 1.78(10)$^{-3}$ \\
-1 & -   & 1.0             & 2.75(10)$^{-7}$ & 2.75(10)$^{-7}$ \\
 0 & -   & -               & 1.0             & 0.99            \\
 1 & -   & -               & -               & 1.0             \\ 

\end{tabular}
\end{table}      

\begin{table}
      \caption{Significance levels computed between the 4 $N$-body
               models for the mass of the percolating cluster 
               statistic.}
\begin{tabular}{|c||cccc|}

Spectral Index & -2 & -1 & 0 & 1 \\
-2 & 1.0 & 3.22(10)$^{-3}$ & 4.51(10)$^{-5}$ & 5.22(10)$^{-8}$ \\
-1 & -   & 1.0             & 1.42(10)$^{-3}$ & 6.91(10)$^{-2}$ \\
 0 & -   & -               & 1.0             & 0.91            \\
 1 & -   & -               & -               & 1.0             \\ 

\end{tabular}
\end{table}      

\section{Conclusions}

In this  paper  we  have  shown  that one can  generate   the standard
percolation statistics from the Minimal Spanning  Tree. This allows us
a large increase in the speed with which  we can perform a percolation
analysis  of  a point dataset.  Our  calculations indicate that we can
gain as much as a factor of 10 in the  computer time needed to perform
the data analysis. This will  become  increasingly important as  large
redshift surveys become available.

In addition we  argue,  based  upon  Poisson distributions, that   the
percolation method is a robust statistical  method when the apropriate
statistic is  used.   Past studies  have argued that   the percolation
threshold as determined   from  the linear extent  of  the percolating
cluster is not a robust  measure of percolation  (Dekel \& West 1985).
We confirm that  result. Contrary to the conclusions  of Dekel \& West
by considering the  behavior of the entire  curve rather than focusing
on a particular parameter of  that curve we find   that a more  robust
estimate of  the percolation properties is  possible.  The mass of the
percolating   cluster appears  to be   very   robust with  respect  to
sampling, as  opposed  to the  linear extent of   the cluster which is
relatively poorly behaved. This is not unexpected  as discussed in the
text  above.  Based upon this  statistic the percolation threshold can
be reliably   estimated even when  the particle  density     varies by
large factors.

We conclude  by  applying these  percolation statistics  to 4 $N$-body
models with different scale-free  Gaussian initial conditions.   Based
upon our comparisons of the curves in figure  4 using the KS test (see
tables 1 \& 2)  it is clear that with  the exception of the  $n=0$ and
$n=1$ models the percolation statistics can easily distinguish between
the models.   Both percolation statistics considered  here are able to
distinguish  between models equally well  (recall a small significance
level indicates that the two distributions are not consistent with the
same parent distribution), but it is only the  mass of the percolating
cluster which is strongly robust  to changes in particle density. Thus
we conclude that percolation may  be a sensitive discriminator between
cosmological models if clustering is not too hierarchical.

\section{Acknowledgments}

One of us (RJS)  would  like to thank   the Center  for  Computational
Sciences  at the    University  of Kentucky for  providing   financial
support.  The  computer  simulations   were performed   on  the Convex
Exemplar at the University of Kentucky. We would also like to thank
Oliver Sk\"oellin and Bill Lahaise for helping with the MST and separtion
code.

%\section{Figure Captions}
%
%\noindent  Figure 1. Percolation  of a  Poisson distribution using the
%percolation code of Bhavsar \& Barrow (1983) in open boxes and the MST
%based algorithm presented  here in filled circles.  The  dimensionless
%separation  length  is  defined  as $l/n^{-1/3}$,  where   $n$  is the
%particle density.
%
%\vspace{0.5cm}
%
%\noindent Figure  2.  Robustness  test  for the linear  extent of  the
%percolating cluster  for  a Poisson  distribution   of particles.  The
%upper left plot is  the entire $64^3$ dataset. The   upper right is  a
%$32^3$ subset, the lower right is a $16^3$  subset, and the lower left
%is  a $8^3$ subset.  The dimensionless  neighborhood radius is defined
%as $l/n^{-1/3}$, where $n$ is the particle density.
%
%\vspace{0.5cm}
%
%\noindent  Figure 3.   Robustness test   for  the  total mass  of  the
%percolating  cluster for  a  Poisson distribution  of particles.   The
%upper  left plot is the entire  $64^3$ dataset. The   upper right is a
%$32^3$ subset, the lower right is a  $16^3$ subset, and the lower left
%is a $8^3$ subset.  The  dimensionless neighborhood radius is  defined
%as $l/n^{-1/3}$, where $n$ is the particle density.
%
%\vspace{0.5cm}
%
%\noindent  Figure 4.  The  percolation statistics applied  to a $32^3$
%subset of particles from a $128^3$ N-body simulation. The 3 statistics
%run    horizontally while  the 4    different  initial conditions  run
%vertically.  The  dimensionless neighborhood radius is  defined
%as $l/n^{-1/3}$, where $n$ is the particle density.

\clearpage
\onecolumn
\begin{figure}
     \epsfxsize = 7.0truein
     \hskip .5truein
     \epsfbox{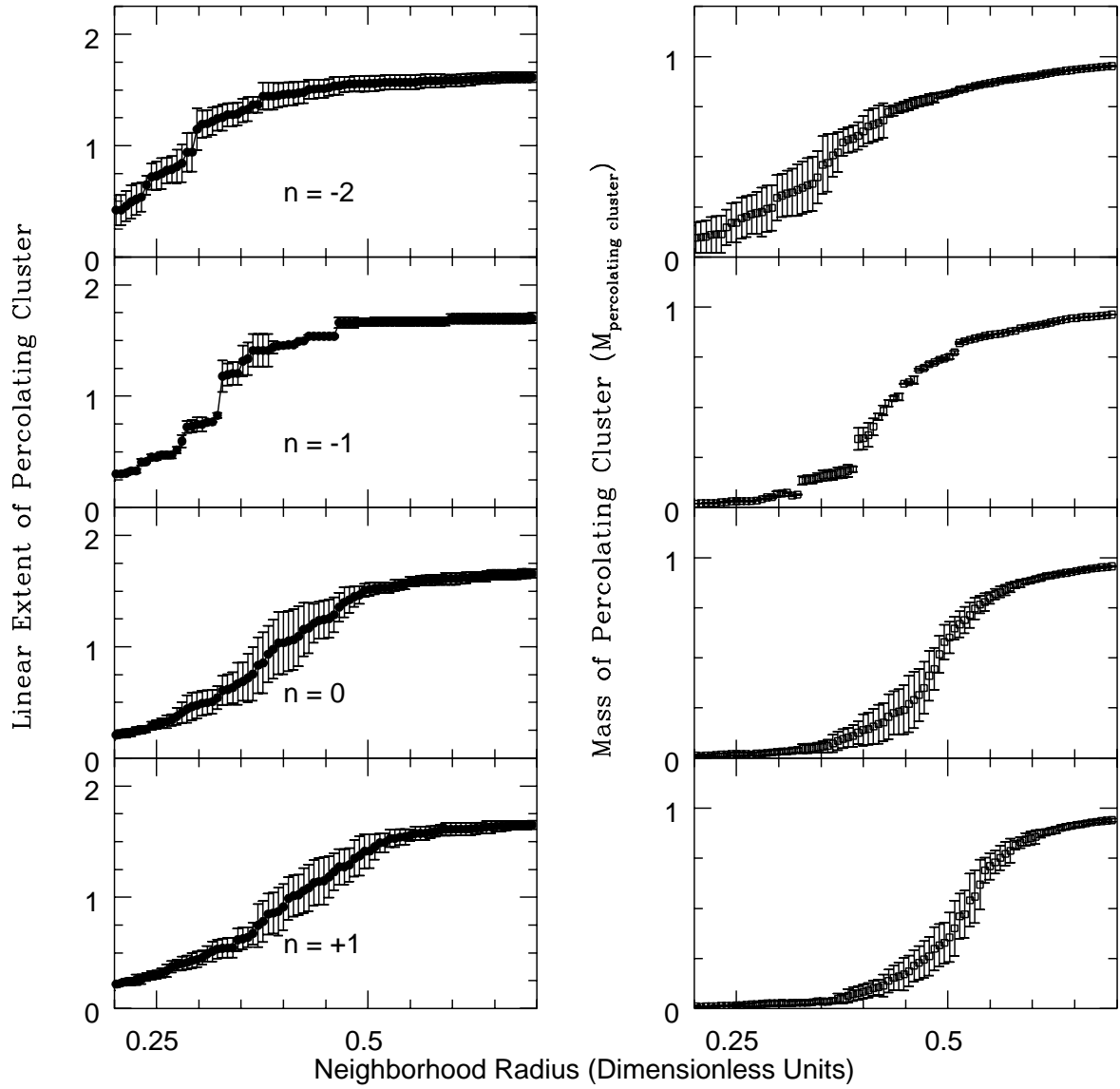}
     \caption{The percolation statistics applied to a $32^3$ subset of
     particles from a $128^3$ N-body simulation. The 3 statistics run
     horizontally while the 4 different initial conditions run vertically.
     The  dimensionless neighborhood radius is  defined
     as $l/n^{-1/3}$, where $n$ is the particle density.}
\end{figure}


\begin{thebibliography}{}

\bibitem[]{ab} 
Abraham, C.  1962, {\it Evaluation of  Clusters on the Basis of 
Random Graph Theory}, IBM Res.  Memo, IBM Corp, York Town Heights, NY

\bibitem[Barrow, et al. 1985]{barrow85}
Barrow, J.D., Bhavsar, S.P., \& Sonoda, D.H. 1985, MNRAS, 216, 17

\bibitem[Bhavsar \& Barrow (1983)]{bhav83}
Bhavsar, S.P., \& Barrow, J. 1983, MNRAS, 205, 61

\bibitem[Bhavsar \& Ling (1988)]{bhav88}
Bhavsar, S.P., \&  Ling, E.N. 1988, ApJ, 331, L63

\bibitem[Davis, et al. (1985)]{davis85}
Davis, M.,  Efstathiou, G.,  Frenk,  C., \&  White, S.D.M. 1985, ApJ,
292, 371

\bibitem[Dekel \& West (1985)]{dekel85}
Dekel, A. \& West, M.J. 1985, ApJ, 288, 411

\bibitem[Einasto, et al (1984)]{einasto84}
Einasto, J., Klypin, A.A., Saar,  E., \& Shandarin, S.F. 1984, MNRAS,
206, 529

\bibitem[Hammersley \& Welsh 1980]{hw}
Hammersley, J.M., \& Welsh, D.J.A. 1980, Contemp Phys, 21, 593

\bibitem[Gower \& Ross (1969]{gow}
Gower, J.C., \& Ross, G.J.S. 1969, Appl Stat, 18, 54

\bibitem[Klypin \& Shandarin 1993]{klypin}
Klypin, A.A., \& Shandarin, S.F. 1993, ApJ, 413, 48

\bibitem[Krzewina \& Saslaw 1995]{krz}
Krzewina, L.G.  \& Saslaw,  W.C.  1995, Preprint,  University    of
Virginia, Department of Astronomy

\bibitem[Melott, et al. 1988]{mel88}
Melott, A.L, Weinberg, D.W., and \& Gott, J.R. 1988, ApJ, 328, 50

\bibitem[Melott 1986]{mel86}
Melott, A.L. 1986, Phys Rev Lett, 56, 1992

\bibitem[Melott \& Shandarin (1983)]{melott83}
Melott, A.L., \& Shandarin, S.F. 1983, ApJ, 410, 469

\bibitem[Press, et al (1992)]{press}
Press, W.H., Flannery, B.P., Teukolsky, S.A., \& Vetterling, W.T. 1992,
{\it Numerical Recipes}, (Cambridge: Cambridge University Press)

\bibitem[Prim 1957]{prim}
Prim, R.C. 1957, Bell Sys Tech J, November, 1389 
 
\bibitem[Shandarin (1983)]{ss83}
Shandarin, S.F. 1983, Soviet Astron. Lett., 9, 104

\bibitem[Shandarin \& Zel'dovich (1983)]{sszel83}
Shandarin, S.F., \& Zel'dovich, Ya.B. 1983, Comments Astrophys., 10, 33

\bibitem[Yess \& Shandarin 1995]{yess}
Yess, C., \& Shandarin, S.F.  1995, astro-ph/9509052 
 
\bibitem[Zahn (1971)]{zahn}
Zahn, C.T. 1971, IEEE Trans Comp, C20, 68

\bibitem[Zel'dovich (1982)]{zel82}
Zel'dovich, Ya.B. 1982, Soviet Astron. Lett., 8, 102


\end{thebibliography}
\end{document}